\documentclass[letterpaper,12pt,single]{article}   %% LaTeX 2e (preferred)
\usepackage{osajnl2} %% do not use with REVTeX4
\usepackage{graphics}
\usepackage{subfigure}
\usepackage{cite}
\usepackage{setspace}
\begin{document}

\title{Analytical solution of the fundamental waveguide mode of $1$D transmission grating for TM polarization}

%% For REVTeX it is possible to automate superscript and e-mail callouts with the superscriptaddress option; see REVTeX4 documentation.

\author{A. T. M. Anishur Rahman$^{*1}$, Krasimir Vasilev$^2$ and Peter Majewski$^1$}
\address{$^1$School of AME, $^2$School of AME and Mawson Institute\\ University of South Australia\\ Mawson Lakes, SA 5095, Australia}
\email{$^*$ Corresponding Author: rahaa001@mymail.unisa.edu.au} %% email address is required

\begin{abstract}
This article presents an analytical solution of the effective index of the fundamental waveguide mode of $1$D metallo-dielectric grating for Transverse Magnetic (TM) polarization. In contrast to the existing numerical solution involving transcendental equation, it is shown that the square of the effective index ($n_{Eff}$) of the fundamental waveguide mode of $1$D grating is inversely proportional to the slit width ($w$) and the refractive index ($n_m$) of the ridge material and varies linearly with the incident wavelength ($\lambda$). Further, it has also been demonstrated that the dependence of $n_{Eff}$ on the grating period ($P$) and the incidence angle ($\theta$) is minimal. Agreement between the results obtained using the solution presented in this article and published data is excellent.
\end{abstract}

% REPLACE WITH CORRECT OCIS CODES FOR YOUR ARTICLE
                          % NOTE: \ocis{} IS ALIASED TO \pacs{} BUT MUST
                          % FORMAT THE TERMS CORRECTLY FOR EACH JOURNAL

\maketitle %% null function with osajnl.sty

\section{Introduction}
\begin{figure}
\centering
\includegraphics{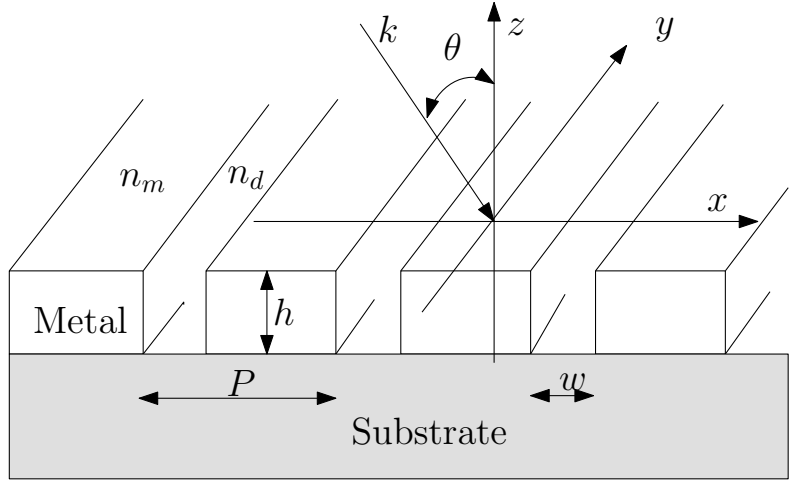}
\centering
\caption{$1$D Lamellar Grating}
\label{fig1}
\end{figure}

 With the advances in nano and micro fabrication technologies, sub-wavelength structures are now readily achievable  \cite{Lalanne2000A,Astilean2000}. Due to their high brightness in resonance, recently $1$D metallo-dielectric grating structures with sub-wavelength slits ($w < \lambda$, see Fig. \ref{fig1}) have been proposed as useful in flat panel displays, Scanning Near-field Optical Microscopy (SNOM), opto-electronic devices, photo-lithography and tunable optical filter \cite{Lalanne2000A,Astilean2000,Pang2007,Kim1999}.

  High brightness or resonance in $1$D grating can be explained using two different theories \cite{Porto1999}. In the regime $\lambda \approx P$, where $\lambda$ is the wavelength of the incident light and $P$ is the period of the grating, coupling between surface plasmon polariton (SPP) of opposite faces of $1$D grating is responsible for enhanced transmission \cite{Porto1999}, whereas for thick enough grating and $\lambda >> P$, resonance coupling between a diffraction order and a waveguide mode plays a major role in the extraordinary transmission through $1$D metallo-dielectric grating structures \cite{Porto1999,Cao2002}. In the latter case, depending upon the incident \textcolor[rgb]{1.00,0.00,0.00}{wavelength}, slit width $w$ and period $P$, different waveguide modes, both propagating and evanescent, are excited inside the slits \cite{Sheng:82}. Propagating modes transfer incident energy from one side of the grating to the other and redistribute the transferred energy among the diffraction orders \cite{Tishchenko:05}. As the slit width $w$ decreases, more and more modes become evanescent and a very few propagating modes survive \cite{Tishchenko:05}. In particular, when $w$ becomes smaller than $\lambda/(2n_d)$, where $n_d$ is the refractive index of the slit/groove, only the fundamental mode propagates and most of the transmitted energy is carried out by this mode \cite{Lalanne2000A,Astilean2000,Porto1999,Takakura2001}. Considering this phenomenon  Lalanne et al. have developed an analytical model of transmission through $1$D grating for TM polarization \cite{Lalanne2000A}. This model can accurately predict resonance wavelengths and their diffraction efficiencies \cite{Cao2002}. Finding transmission efficiency using this model requires effective index of the fundamental mode, which is defined as $n_{Eff}=k_z/k_0$, where $k_z$ and $k_0$ are \textcolor[rgb]{1.00,0.00,0.00}{the $z$-component of the wave vector} of the fundamental waveguide mode and \textcolor[rgb]{1.00,0.00,0.00}{the  wave number of the free space} incident electromagnetic illumination respectively \cite{Astilean2000}. This model also depends on the grating parameters i.e. $w$, $P$ and $h$, where $h$ is the thickness of the gratings. Similarly, Porto et al. \cite{Porto1999} and Garcia-Vidal et al. \cite{GarciaVidal2002} have developed models of transmission through $1$D gratings by considering only the fundamental mode and their results agree closely with those of Lalanne et al. \cite{Lalanne2000A}. Profile of the fundamental mode and $x-$ and $z-$ components of its wave \textcolor[rgb]{1.00,0.00,0.00}{vector} ($k_x$ and $k_z$, see Fig. \ref{fig1}) as well as those of other modes can be determined numerically by solving transcendental modal equation proposed by Sheng et al. in $1982$ \cite{Sheng:82}. This method is known as modal analysis and its solutions, also known as eigenmodes, correspond to various modes of the waveguide structure. Overall this method provides exact description of the modes \cite{Gaylord1985,Tishchenko:05} and is becoming popular \cite{Clausnitzer2005,Catchpole2007,Lyndin:07,Foresti:06} due to its phenomenological interpretation of the wave propagation via grating structure. Despite this, the method is still a numerical technique and, as inherent to numerical techniques, is devoid of physical insights i.e. can not provide a direct relationship (such as $n_{Eff}^2$ varies inversely with $w$) among $n_{Eff}$, $w$, $P$, $\lambda$, $\theta$ (incidence angle) and $n_m$ (refractive index of the ridge metal) and hence the physics behind the wave propagation via $1$D gratings is not well understood. Also, finding a solution of the transcendental equation requires searching inside the variable domains \cite{Foresti:06} and is computationally demanding \cite{Gaylord1985}.

In contrast to the existing numerical solution, in this article we attempt to provide an explicit relation involving $n_{Eff}$, $w$, $P$, $n_m$, $\theta$ and $\lambda$. This direct relationship between $n_{Eff}$ and the grating parameters provides a vivid explanation of the physics behind the wave propagation via $1$D grating structures. To the best our knowledge, this kind of analytical model relating $n_{Eff}$ and the grating parameters is nonexistent in the literature for $1$D waveguide structures even though such a relationship exists for $2$D waveguides \cite{Feynman2006}. Further, as with analytical solutions, finding $n_{Eff}$ using the method presented here is easy and does not require searching inside the variable domains and consequently it is computationally much less demanding. Also, results obtained using our model agree very closely to those of the exact numerical calculation. In the process of deriving our main result, we assume $w < \lambda/(2n_d)$.

\section{Analytical model}
%%%\section{Analytical solution of the fundamental TM mode of $1$D grating}
Let us consider a TM polarized electromagnetic wave $H_y=\exp{(ik_0(\sin\theta x-\cos\theta z))}* \exp{(-i\omega t)}$ is incident upon the metallo-dielectric grating of Fig. \ref{fig1} at an incidence angle $\theta$. This incident wave excites various waveguide modes which in turn transfer energy from the incident side ($z>0$, see Fig. \ref{fig1}) of the grating structure to the outgoing side ($z < -h$). \textcolor[rgb]{1.00,0.00,0.00}{$x-$ and $z-$ components of wave vectors} corresponding to different waveguide modes can be obtained by solving transcendental Eq. (\ref{eq1}) \cite{Sheng:82}.

\begin{eqnarray}
\cos(k_0P\sin\theta)-\cos(\beta rP)\cos(\alpha
fP)+
\frac{1}{2}[\frac{\epsilon_m\alpha}{\beta}+
\frac{\beta}{\epsilon_m\alpha}]\sin(\beta
rP)\sin(\alpha fP)=0
\label{eq1}
\end{eqnarray}
where $\alpha=k_0\sqrt{\epsilon_d-n_{Eff}^2}$ and $\beta=k_0\sqrt{\epsilon_m-n_{Eff}^2}$ are the $x-$ components of a waveguide mode in the slit/groove and ridge material respectively. $r=(P-w)/P$, $f=1-r=w/P$, $\epsilon_d=n_d^2$ is the dielectric constant of the slit and $\epsilon_m=n_m^2$ is the dielectric constant of the grating ridge. For metallic ridges, dielectric constant is given by $\epsilon_m=n_m^2=(\eta+i\kappa)^2$, where $\eta$ and $\kappa$ are the real and imaginary components of the refractive index.  When $|\epsilon_m| >> |n_{Eff}^2|$, $\beta$ can be written as $\beta=k_0n_m$. Assuming $\epsilon_d=n_d^2=1$, $\alpha$ can be written as $\alpha=k_0\rho$, where $\rho=\sqrt{1-n_{Eff}^2}$. Considering above Eq. (\ref{eq1}) can be
rewritten as Eq. (\ref{eq2}).

\begin{eqnarray}
\cos(k_0P\sin\theta)-\cos(k_0n_mrP)\cos(k_0\rho
fP)+
\frac{1}{2}[\frac{\epsilon_m\alpha}{\beta}+\frac{\beta}{\epsilon_m\alpha}]\sin(k_0n_m
rP)\sin(k_0\rho fP)=0
\label{eq2}
\end{eqnarray}

Given that $|\epsilon_m| >> |n_{Eff}^2|$ and $|\epsilon_m\rho^2| >> 1$, the first factor of the $3^{rd}$ term
of Eq. (\ref{eq2}) can be written as $\frac{1}{2}[\frac{k_0\rho\epsilon_m}{k_0\sqrt\epsilon_m}+\frac{k_0\sqrt{\epsilon_m}}{k_0\rho \epsilon_m}]$ $\approx\sqrt{\epsilon_m}\rho/2=n_m\rho/2$. Based upon this Eq. (\ref{eq2}) can be written as Eq. (\ref{eq3}).

\begin{eqnarray}
\cos{(k_0P\sin{\theta})}-\cos{q}\cos{p}+\frac{1}{2}n_m\rho\sin{q}\sin{p}=0
\label{eq3}
\end{eqnarray}
where $p=2\pi \rho w/\lambda$ and $q=2\pi n_m (P-w)/\lambda$. Expanding $\sin p$ and $\cos p$ into Taylor series, Eq.
(\ref{eq3}) can be expressed as Eq. (\ref{eq4}).

\begin{eqnarray}
\nonumber
&&\frac{1}{2}n_m\rho
(p-\frac{p^3}{3!}+\frac{p^5}{5!}-...)\sin q-(1-\frac{p^2}{2!}
+\frac{p^4}{4!}-\frac{p^6}{6!}+...)\cos
q+\cos(k_0P\sin\theta)=0 \\
%\nonumber
&&\frac{n_m}{4\pi f\gamma}
(p^2-\frac{p^4}{3!}+\frac{p^6}{5!}-...)\sin q
-(1-\frac{p^2}{2!}+\frac{p^4}{4!}-\frac{p^6}{6!}+...)\cos
q+\cos(k_0P\sin\theta)=0\label{eq4}
\end{eqnarray}

In the typical operating conditions where only the fundamental mode survives such as ($r=0.8571$, $f=0.1429$, $|n_m| \approx 18.73$)\cite{Porto1999} , ($0.90\le r \le 0.9889$, $0.0111 \le f \le 0.10$, $|n_m|\approx 6.796$)\cite{Astilean2000}  and ($0.50\le r \le 0.95$, $0.05 \le f \le 0.50$, $|n_m| \approx 5.03$)\cite{Foresti:06}, $|q|$ becomes greater than $1$ and $|p|$ is less than unity. In this case any power of $p$ above $4$ in Eq. (\ref{eq4}) can be neglected. After some simple algebraic manipulations, one can write Eq. (\ref{eq4}) as Eq. (\ref{eq5}).

\begin{eqnarray}
Ap^4-Bp^2+C=0
\label{eq5}
\end{eqnarray}

where $A=(\frac{n_m}{\pi f\gamma}\frac{\sin q}{4!}+\frac{\cos q}{4!})$, $B=(\frac{n_m}{2\pi f\gamma}\frac{\sin q}{2!}+\frac{\cos q}{2!})$, $C=(\cos q-\cos(k_0P\sin\theta))$. Eq. (\ref{eq5}) can be easily solved using the standard algebra and the solutions are given in Eq. (\ref{eq6}).

\begin{eqnarray}
\nonumber
p^2&=&\frac{B\pm\sqrt{B^2-4AC}}{2A}\\
n_{Eff}^2&=&1-\frac{3\lambda^2}{4\pi^2w^2}[1+\frac{1}{D}\pm\sqrt{1-[\frac{2\cos q-8\cos {(k_0P\sin \theta)}}{3\cos q}]\frac{1}{D}+\frac{1}{D^2}}]
\label{eq6}
\end{eqnarray}

where $D=1+\frac{\lambda n_m}{\pi w}\frac{\sin q}{\cos q}$. $\sin q$ and $\cos q$ can be expanded as $\sin{\{k_0(P-w)\eta\}}*\cosh{\{k_0(P-w)\kappa\}} + i \cos{\{k_0(P-w)\eta\}}\sinh{\{k_0(P-w)\kappa\}}$ and $\cos{\{k_0(P-w)\eta\}}*$ $\cosh{\{k_0 (P-w) \kappa\}}$ $-i \sin{\{k_0(P-w)\eta\}}*\sinh{\{k_0(P-w)\kappa\}}$ respectively. For $k_0(P-w)\kappa > 1$, $\cosh{\{k_0(P-w)\kappa\}}\approx \sinh{\{k_0(P-w)\kappa\}}$. In this case, $\sin{q}$ and $\cos{q}$ can be written as - $\cosh{\{k_0(P-w)\kappa\}} * \exp{[i\{\pi/2-k_0(P-w)\eta\}]}$ and $\cosh{\{k_0(P-w)\kappa\}}\exp{[ik_0(P-w)\eta]}$ respectively. After some simple manipulation one can find Eq. (\ref{eq7}).

\begin{eqnarray}
\nonumber
n_{Eff}^2=1-\frac{3\lambda^2}{4\pi^2w^2}[1+\frac{\pi w}{\pi w+i\lambda n_m}\pm [1-[\frac{2}{3}-\frac{8\cos {(k_0P\sin \theta)}\exp{\{ik_0(P-w)\eta\}}}{3\cosh{\{k_0(P-w)\kappa\}}}] \\\frac{\pi w}{\pi w+i\lambda n_m}+\frac{\pi^2 w^2}{(\pi w+i\lambda n_m)^2}]^{1/2}]
\label{eq7}
\end{eqnarray}

Given that $\cosh{\{k_0(P-w)\kappa\}} >> 1$ and $|\frac{\pi w}{\pi w+i\lambda \eta}|< 1$, the term under the square root in Eq. (\ref{eq7}) can be expanded into binomial series. Neglecting terms of the expansion with power two or more, Eq. (\ref{eq7}) can be expressed as Eq. (\ref{eq8}).

\begin{eqnarray}
\nonumber
n_{Eff}^2=1-\frac{3\lambda^2}{4\pi^2w^2}[1+\frac{\pi w}{\pi w+i\lambda n_m}\pm [1-[\frac{1}{3}-\frac{4\cos{(k_0 P\sin{\theta})}\exp{\{ik_0(P-w)\eta\}}}{3\cosh{\{k_0(P-w)\kappa\}}}]\\\frac{\pi w}{\pi w+i\lambda n_m}+\frac{\pi^2 w^2}{2(\pi w+i\lambda n_m)^2}]]
\label{eq8}
\end{eqnarray}

Provided that $Re(n_{Eff})\ge 1$, the solution corresponding to the fundamental mode can be written as Eq. (\ref{eq9}).
\begin{eqnarray}
\nonumber
n_{Eff}^2=&1-[1-\frac{\cos{(k_0 P\sin{\theta})}\exp{\{ik_0(P-w)\eta\}}}{\cosh{\{k_0(P-w)\kappa\}}}]\frac{\lambda^2}{\pi w(\pi w+i\lambda n_m)}+\frac{3\lambda^2}{8(\pi w+i\lambda n_m)^2}\\
=&1+i[1-\frac{\cos{(k_0 P\sin{\theta})}\exp{\{ik_0(P-w)\eta\}}}{\cosh{\{k_0(P-w)\kappa\}}}]\frac{\lambda}{\pi n_m w(1-i\frac{\pi w}{\lambda n_m})}-\frac{3}{8n_m^2(1-i\frac{\pi w}{\lambda n_m})^2}
\label{eq9}
\end{eqnarray}
Considering $0 < |\frac{i\pi w}{\lambda n_m}|< 1$ and expanding the denominators of the $2^{nd}$ and $3^{rd}$ terms of Eq. (\ref{eq9}) into binomial series of $\frac{i\pi w}{\lambda n_m}$ and keeping only the first and second power of $w$ and $n_m$ of the expansion respectively, one can represent Eq. (\ref{eq9}) as Eq. (\ref{eq10}).

\begin{eqnarray}
n_{Eff}^2=1-[\frac{11}{8n_m^2}-i\frac{\lambda}{\pi w n_m}]+[\frac{1}{n_m^2}-i\frac{\lambda}{\pi wn_m}] \frac{\cos{(k_0 P\sin{\theta})}}{\cosh{\{k_0(P-w)\kappa\}}}\textcolor[rgb]{1.00,0.00,0.00}{\exp}\{ik_0(P-w)\eta\}
\label{eq10}
\end{eqnarray}

\section{Results and discussions}
\begin{figure}
\doublespacing
\centering
\includegraphics[width=7.5cm]{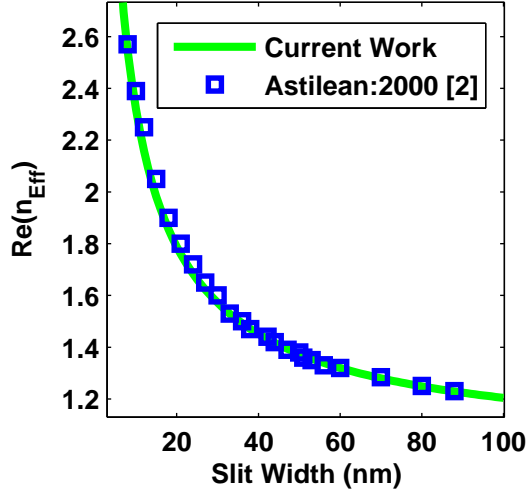}
\centering \caption{Real part of the effective index of the fundamental mode as a function of slit width
corresponding to $n_d=1$, $\theta=0^o$, $P=900$ nm, $\lambda=1433$ nm for silver gratings}
\label{fig2}
\end{figure}

The real component of $n_{Eff}$ corresponding to the fundamental mode obtained from Eq. (\ref{eq10}) is plotted in Fig. \ref{fig2} for $n_d=1$, $\theta=0^o$, $P=900$
nm and $\lambda =1433$ nm for silver grating as a function of slit width $w$. Grating parameters used in this example have been taken from Astilean et al. \cite{Astilean2000}, where the authors show how the effective index of the fundamental mode evolves as the slit width of the grating changes. For the purpose of comparison $Re(n_{Eff})$ from Ref.\cite{Astilean2000} has also been included in Fig. \ref{fig2}. One can see that there is an excellent agreement between our results and those from Ref.\cite{Astilean2000}. It is also evident that as $w$ increases, $Re(n_{Eff})$ decreases or there is an inverse relationship between $Re(n_{Eff})$ and $w$. To confirm this let us look more closely at Eq. (\ref{eq10}). Upon consideration one can find that the contribution from the $3^{rd}$ term in Eq. (\ref{eq10}) toward $n_{Eff}^2$ and hence toward $Re(n_{Eff})$ is very negligible since $\cosh{\{k_0(P-w)\kappa\}} >> 1$, $|\exp\{ik_0(P-w)\eta\}| \le 1$ and $|\cos{(k_0P\sin{\theta})}| \le 1$. Considering this one can rewrite Eq. (\ref{eq10}) as Eq. (\ref{eq11}) from which it is understandable that for fixed $P$, $n_m$, $\lambda$ and $\theta$, $n_{Eff}^2$ and therefore $n_{Eff}$ vary inversely with $w$. At this point we quickly mention that this kind of physical insight is not understandable from the existing numerical solution.

\begin{eqnarray}
n_{Eff}^2=1-[\frac{11}{8n_m^2}-i\frac{\lambda}{\pi w n_m}]
\label{eq11}
\end{eqnarray}

\begin{figure}
\centering
\doublespacing
\includegraphics[width=7.5cm]{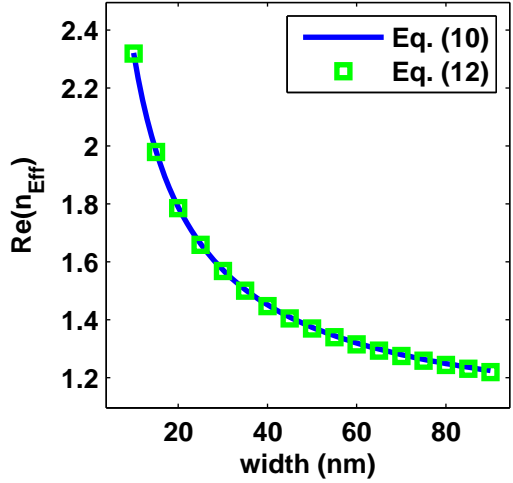}
\centering \caption{Real part of the effective index of the fundamental mode as a function of slit width
corresponding to $n_d=1$, $\theta=0^o$, $P=900$ nm, $\lambda=1433$ nm for silver gratings}
\label{fig3}
\end{figure}

From Eq. (\ref{eq11}), one can also find that as $w \rightarrow 0$,  $n_{Eff}^2$ approaches infinity as the grating becomes impermeable to light. On the other hand, when $w \rightarrow \lambda/2$, $n_{Eff}^2$ approaches a constant value of $1+\delta$, where $\delta=i\frac{\lambda}{\pi w n_m}-\frac{11}{8n_m^2}$ depends on the dielectric constant of the ridge material and should be much much smaller than unity as $|n_m^2| >> 1$ has been assumed. When $w=\lambda/2$ and $|n_m^2| >> 1$ (which is true for most of the metals in the infrared region of the electromagnetic spectrum), $\delta \rightarrow 0$ and $n_{Eff}^2$ (accordingly $n_{Eff}$) approaches unity as expected. However, when $0 < w < \lambda/2$ and $|n_m^2| >> 1$, Eq. (\ref{eq11}) can be further simplified to Eq. (\ref{eq12}) from which one can observe that $n_{Eff}^2$ is inversely proportional to $w$ and $n_m$ and varies linearly with $\lambda$. For the purpose of demonstration we have plotted data corresponding to Eq. (\ref{eq10}) and Eq. (\ref{eq12}) in Fig. \ref{fig3} for the same set of grating parameters of Ref.\cite{Astilean2000}. It can be seen from Fig. \ref{fig3} that there is no difference between the curves corresponding to Eq. (\ref{eq10}) and Eq. (\ref{eq12}).

\begin{eqnarray}
n_{Eff}^2=1+i\frac{\lambda}{\pi w n_m}
\label{eq12}
\end{eqnarray}

\begin{figure}
\doublespacing
\centering
\subfigure{
\includegraphics[width=7.5cm]{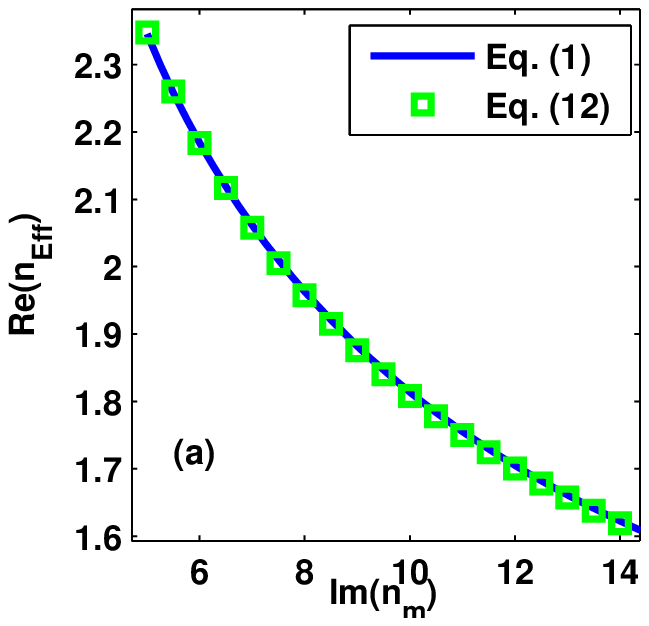}}
\subfigure{
\includegraphics[width=7.5cm]{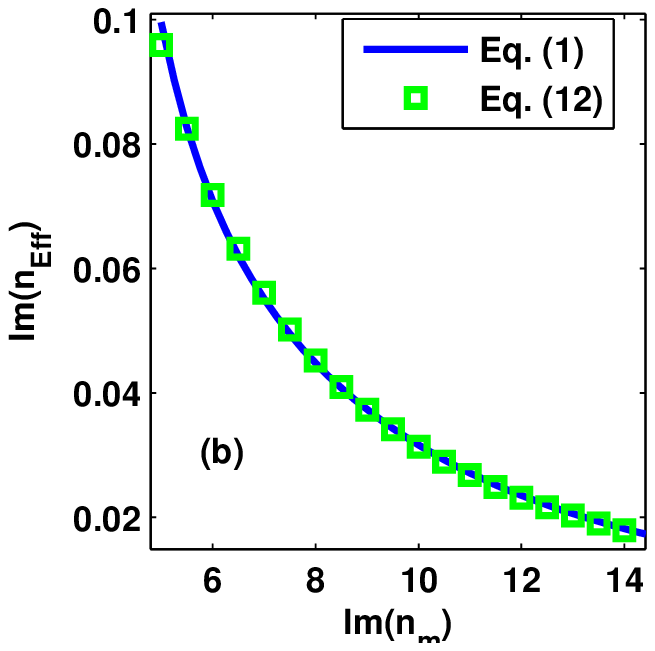}}
\centering \caption{(a) real and (b) imaginary components of the effective index of the fundamental mode as a function of $Im(n_m)$ while keeping $Re(n_m)$ constant corresponding to $n_d=1$, $\theta=0^o$, $w=21$ nm, $P=150$ nm, $\lambda=1500$ nm. Grating parameters have been taken from Ref.\cite{Rahman2011A} }
\label{fig4}
\end{figure}

\begin{figure}
\centering
\doublespacing
\subfigure{
\includegraphics[width=7.5cm]{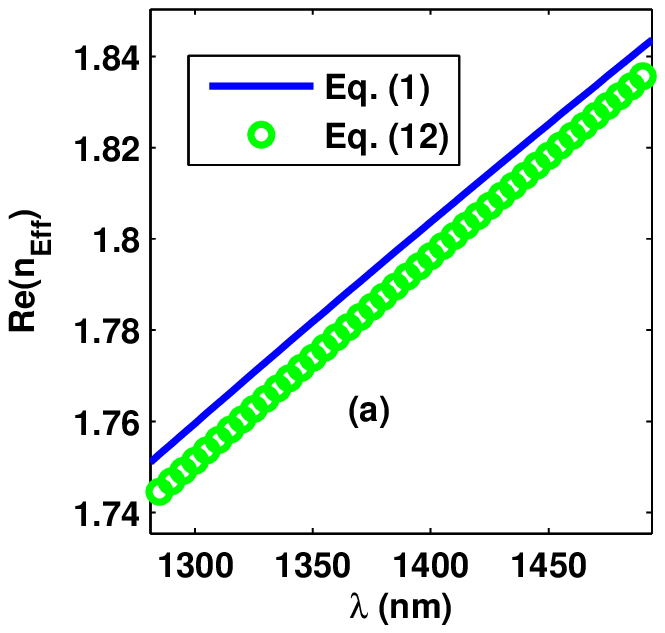}}
\subfigure{
\includegraphics[width=7.5cm]{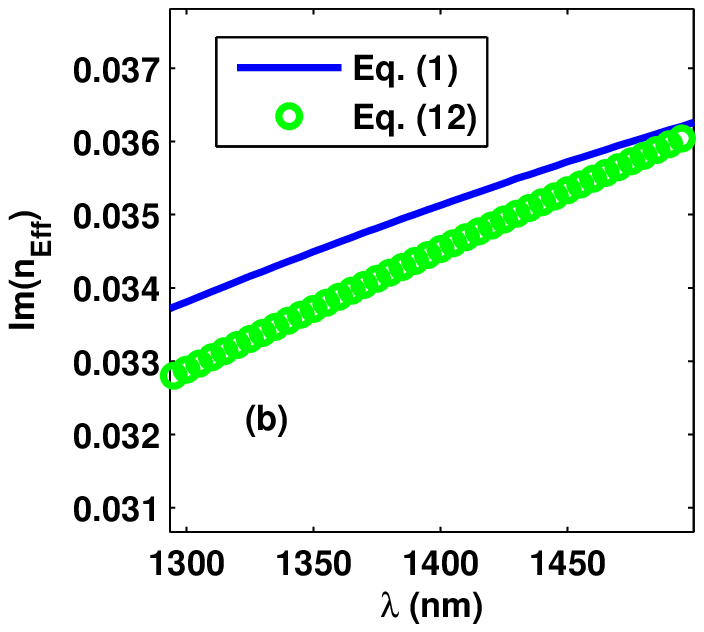}}
\centering \caption{(a) real and (b) imaginary components of the effective index of the fundamental mode as a function of $\lambda$ where $n_d=1$, $\theta=0^o$, $w=21$ nm and $P=150$ nm. $n_m=0.530+9.5070i$ is that of gold at $\lambda=1500$ nm. Grating parameters have been taken from Ref.\cite{Rahman2011A} }
\label{fig4a}
\end{figure}

Moreover to verify that $n_{Eff}^2$ varies inversely with $n_m$, in Fig. \ref{fig4} we have plotted both $Re(n_{Eff})$ and $Im(n_{Eff})$ as a function of the refractive index of the ridge material using the exact transcendental Eq. (\ref{eq1}) and Eq. (\ref{eq12}). This is equivalent to considering various ridge materials while keeping geometrical grating parameters i.e. $w$, $P$ and $h$ intact. Grating parameters have been taken from \cite{Rahman2011A}, where designing $1$D grating for extraordinary optical transmission is considered using the numerical optimization technique. From Fig. \ref{fig4}, one can see that there is an excellent agreement between the exact method and the simple analytical method we have presented above. It is also noticeable that as $Im(n_m)$ increases, loss associated with the fundamental mode ($Im(n_{Eff})$) decreases. This is due to fact that as the imaginary component of the dielectric constant increases, a metal becomes highly reflective and waves can propagate without incurring much loss. Further, although $\lambda$ and $n_m$ are related, for completeness in Fig. \ref{fig4a} we have plotted both $Re(n_{Eff})$ and $Im(n_{Eff})$ as a function of $\lambda$ while keeping $n_m$, $w$, $P$ and $h$ constant for $\theta=0^o$. The grating geometrical parameters have been taken from Ref. \cite{Rahman2011A} like before and $n_m=0.530+9.5070i$ is that of gold at $1500$ nm. Data for this graph have been obtained from the exact numerical calculation (Eq. (\ref{eq1})) and Eq. (\ref{eq12}). It can be observed that $Re(n_{Eff})$ and $Im(n_{Eff})$ vary linearly with $\lambda$ as predicted by Eq. (\ref{eq12}) and the agreement between the exact calculation and the simplistic model of $n_{Eff}$ presented in Eq. (\ref{eq12}) is very good.

\begin{figure}
\centering
\doublespacing
\subfigure{
\includegraphics[width=7.5cm]{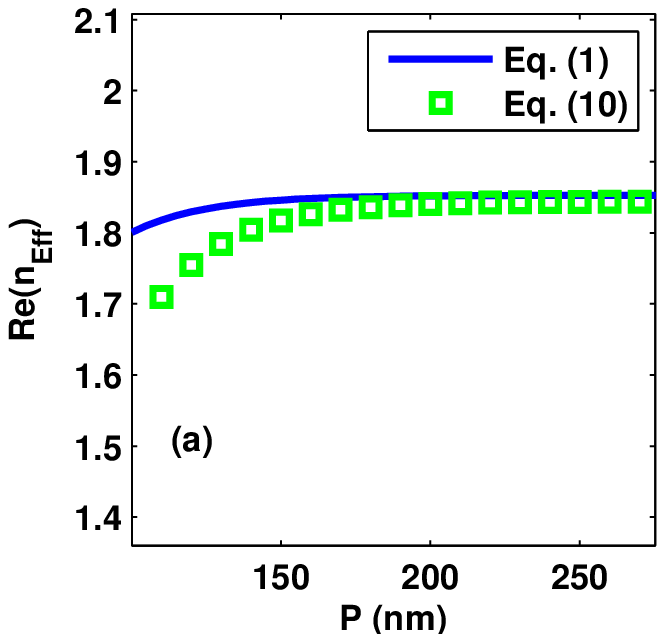}}
\subfigure{
\includegraphics[width=7.5cm]{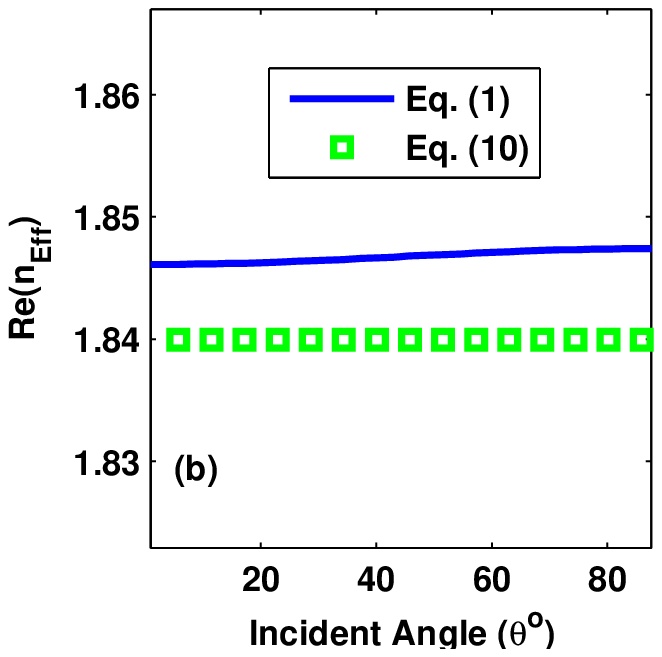}}
\centering \caption{$Re(n_{Eff})$ as a function of (a) $P$ and (b) incidence angle $\theta$ corresponding to $n_d=1$, $w=21$ nm and $\lambda=1500$ nm. For (b) $P$ is equal to $150$ nm.}
\label{fig5}
\end{figure}

 To complete the investigation of the dependence of $n_{Eff}$ on the grating parameters, let us consider the impact of the two remaining parameters, namely the grating period and the incidence angle on $n_{Eff}$. According to Eq. (\ref{eq12}), $n_{Eff}^2$ and consequently $n_{Eff}$ do not depend upon these two parameters. However, if one considers Eq. (\ref{eq10}), then it is found that the dependence of $n_{Eff}^2$ on $P$ and $\theta$ is very weak. To confirm this we have plotted $n_{Eff}$ as a function of $P$ and $\theta$ in Fig. \ref{fig5} (a) and (b) respectively. In both cases there is a very good qualitative agreement between the exact result obtained by numerically solving Eq. (\ref{eq1}) and the calculation performed using Eq. (\ref{eq10}). It can be observed that the dependence of $n_{Eff}$ on $P$ and $\theta$ is very minimal. In particular, the difference between $Re(n_{Eff})$ corresponding to the two extreme incidence angles ($\theta=0^o$ and $\theta=90^o$) is approximately $0.076\%$. This is due to the fact that the denominator of the $3^{rd}$ term of Eq. (\ref{eq10}) is much larger than the $\theta$ dependent $\cos{(k_0P\sin{\theta})}$ (varies between $-1$ and $+1$) in the numerator. Consequently, a relatively small variation in $\cos{(k_0P\sin{\theta})}$ caused by the variation in $\theta$ does not resonate a significant change in $n_{Eff}$. On the other hand, the dependence of $n_{Eff}$ on $P$ is discernable up to a certain value of the grating period, beyond that it becomes independent of P. This behavior of $n_{Eff}$ in regards to $P$ can be explained by considering Eq. (\ref{eq10}) again.  For a fixed $w$, $n_m$ and $\theta$, $\cosh{\{k_0(P-w)\kappa\}}$ is a real number and is greater than $1$. As $P$ and hence $(P-w)$ increases, the denominator of Eq. (\ref{eq10}) gets bigger and bigger. When the value of $\cosh{\{k_0(P-w)\kappa\}}$ is relatively small, the contribution from the $3^{rd}$ of Eq. (\ref{eq10}) towards $n_{Eff}^2$ is appreciable but when it becomes immensely large, the $3^{rd}$ term from Eq. (\ref{eq10}) can be completely ignored and $n_{Eff}^2$ becomes independent of $P$ and approaches the value predicted by Eq. (\ref{eq12}). In general, it can be concluded that if $2\pi(P-w)\kappa > 10\lambda$ then $n_{Eff}$ does not depend on $P$ significantly. Further, considering the above discussion, one can conclude that the analytical model presented in Eq. (\ref{eq12}) for the fundamental mode of $1$D grating structure is a very good representation of the exact solution and captures all the physics of the fundamental mode propagation via $1$D grating structure. Another interesting point that can observed from Eq. (\ref{eq12}) is that, unlike the $2$D structures such as the rectangular waveguides \cite{Feynman2006}, $1$D waveguide has no cutoff wavelength above which all modes including the fundamental waveguide mode are non-propagating.

\begin{figure}
\centering
\doublespacing
\includegraphics[width=7.5cm]{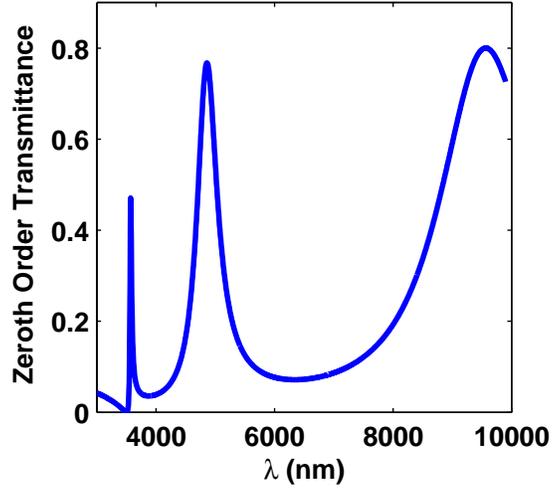}
\centering \caption{Zeroth order transmittance corresponding to $P=3.50$ $\mu$m, $w=0.50$ $\mu$m, $h=4.00$ $\mu$m, $n_d=1$ and $\theta=0^o$ \cite{Cao2002}. $n_m$ is that of gold \cite{Palik85}. Transmission efficiency is based on the model of Ref. \cite{Lalanne2000A, Cao2002} where $n_{Eff}$ is needed to complete the calculation. In Ref. \cite{Cao2002} $n_{Eff}$ has been found using a technique called method of line while in plotting this graph we have used Eq. (\ref{eq12})}.
\label{fig6}
\end{figure}

Additionally, to show that the analytical solution of \textcolor[rgb]{1.00,0.00,0.00}{$n_{Eff}$} presented in this article is suitable for different \textcolor[rgb]{1.00,0.00,0.00}{ridge materials, incidence angles and geometrical grating parameters}, we present more data from the literature and compare them with the data generated using the model presented in Eq. (\ref{eq12}). In this regard, we consider data from Ref. \cite{Foresti:06} where the authors consider aluminum gratings, from Ref. \cite{Astilean2000} where silver grating is investigated and from Ref. \cite{Lyndin:07} in which loss less metals are considered. Comparative results are presented in Table \ref{tab1}. In all cases, irrespective of ridge materials and incidence angles, the agreement between the current results and those from the literature is good. We present one more example from Ref. \cite{Cao2002} where the authors show the negative roles of SPP on EOT for the case of $1$D transmission grating for the TM case. In their analysis the authors find the optical transmission of the zeroth diffraction order via $1$D grating by using a one-mode (fundamental mode) model of optical transmission \cite{Lalanne2000A}. In this model effective index of the fundamental mode is necessary and is determined by the method of line \cite{Lalanne2000A}. As per the authors analysis, three different transmission peaks appear at $3.58$ $\mu$m, $4.9$ $\mu$m and $9.5$ $\mu$m corresponding to a gold grating with $w=0.50$ $\mu$m, $h=4.00$ $\mu$m, $P=3.50$ $\mu$m, $n_d=1$ and $\theta=0^o$. For the purpose of comparison we have plotted the zeroth order transmission efficiency using the same set up of Ref. \cite{Cao2002} except $n_{Eff}$ which we have determined using the solution developed in this article (Eq. (\ref{eq12})). From Fig. \ref{fig6} it can be seen that there are three extraordinary optical transmission peaks at $3.57$ $\mu$m, $4.9$ $\mu$m and $9.56$ $\mu$m. Upon comparison with the data from Ref. \cite{Cao2002}, one can find that the agreement between the data generated using our analytical solution and those by calculating $n_{Eff}$ numerically is very good.

\begin{table}
\centering
\doublespacing
\caption{Effective index corresponding to the fundamental mode with $n_d=1$. $\lambda$, $w$ and $P$ are given in {nm}}
\small
\begin{tabular}{|c|c|c|} % centered columns (4 columns)
\hline
\hline
Grating Parameters & $n_{Eff}$& $n_{Eff}$\\
&Literature&Present Work\\
\hline
Ridge- Silver&&\\
$\lambda=1183$, $w=90$&&\\$P=900$, $\theta=0^o$&$1.224+0.002i$ \cite{Astilean2000}&$1.220+0.002i$\\
\hline
Ridge- Unknown&&\\
$\lambda=632.8$, $w=93.52$&&\\ $P=500$, $\theta=30^o$&$1.105$ \cite{Lyndin:07}&$1.103$\\
\hline
Ridge- Aluminum&&\\
$\lambda=450$, $w=100$&&\\ $P=200$, $\theta=35^o$&$1.142+0.015i$ \cite{Foresti:06}&$1.133+0.013i$\\
\hline
\hline
\end{tabular}
\label{tab1}
\end{table}

Finally, we stress that the accuracy of the solution of the fundamental mode presented above significantly depends on the validity of the assumption $|n_{Eff}^2| << |\epsilon|$ and whenever this condition is not satisfied there will be a mismatch between the $n_{Eff}$'s calculated by \textcolor[rgb]{1.00,0.00,0.00}{Eq. (\ref{eq1})} and Eq. (\ref{eq12}). It is also important to mention that retaining higher powers of $p$ in Eq. (\ref{eq5}) does not significantly improve the accuracy of $n_{Eff}$ but increases processing difficulties and overall $n_{Eff}$ becomes an obscure function of $w$, $P$, $\theta$, $\lambda$ and $n_m$. Lastly, even though we have not considered other grating modes explicitly, conclusions similar to those of the fundamental mode may be applicable to them.

\section{Conclusion}
In conclusion, we have provided a simple analytical solution of the effective index of the fundamental waveguide mode of $1$D grating structure for TM polarization. It has been shown that the square of the effective index of the fundamental waveguide mode is inversely proportional to the slit width and the refractive index of the ridge material. Dependence of $n_{Eff}$ on the grating period and the incidence angle is negligible. The solution provided in this work is very easy to compute and produces results that match closely to those of the exact method. We have also demonstrated that irrespective of the grating materials, incidence angles and incidence wavelength, the analytical solution presented in this article provides reliable results.

\end{document}